\documentclass{article}

\usepackage{xcolor}
\usepackage{graphicx}
\usepackage{url}
\usepackage{footnote}

\begin{document}

\title{CoVoL: A Cooperative Vocabulary Learning Game for Children with Autism}

\author{Pawel Chodkiewicz, Pragya Verma, Grischa Liebel, \\ 
Department of Computer Science, Reykjavik University, Iceland\\\{pawel23,pragyav,grischal\}@ru.is}


\date{This is a pre-print version of the paper with the same title published in the 24th annual ACM Interaction Design and Children (IDC) Conference.}
\maketitle

\begin{abstract}
Children with Autism commonly face difficulties in vocabulary acquisition, which can have an impact on their social communication. Using digital tools for vocabulary learning can prove beneficial for these children, as they can provide a predictable environment and effective individualized feedback. While existing work has explored the use of technology-assisted vocabulary learning for children with Autism, no study has incorporated turn-taking to facilitate learning and use of vocabulary similar to that used in real-world social contexts.
To address this gap, we propose the design of a cooperative two-player vocabulary learning game, CoVoL. CoVoL allows children to engage in game-based vocabulary learning useful for real-world social communication scenarios. 
We discuss our first prototype and its evaluation.
Additionally, we present planned features which are based on feedback obtained through ten interviews with researchers and therapists, as well as an evaluation plan for the final release of CoVoL.
\end{abstract}




\section{Introduction}
\begin{figure}
    \centering
    \begin{minipage}{0.45\textwidth}
        \centering
        \includegraphics[width=0.75\textwidth]{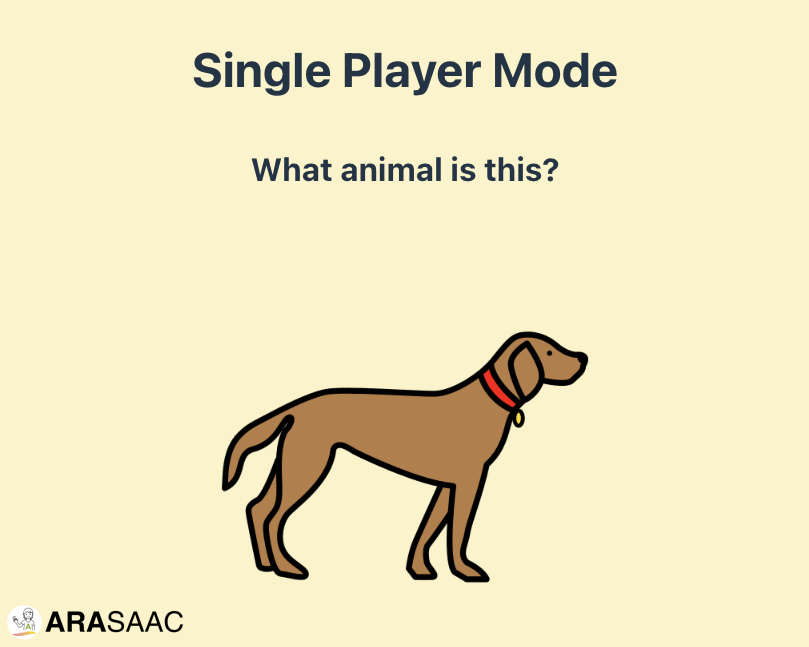}
        \caption{Example prompt for a dog.}
        \label{fig:promptExample}
    \end{minipage}\hfill
    \begin{minipage}{0.45\textwidth}
        \centering
        \includegraphics[width=0.8\textwidth]{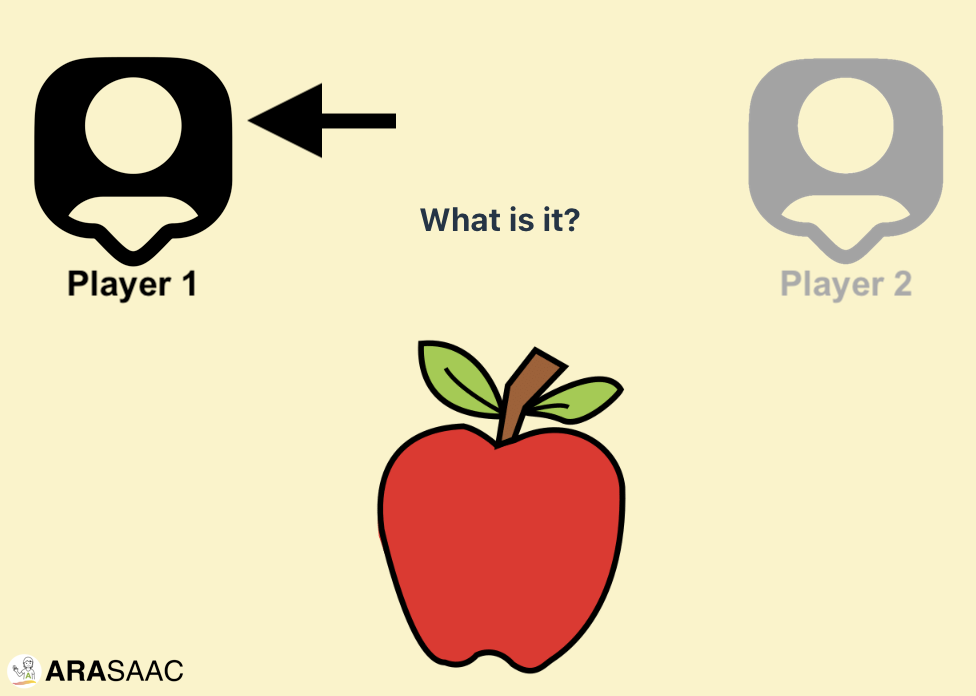}
        \caption{Example prompt for an apple in two-player mode. It is Player 1's turn.}
  \label{fig:promptExampleTwoPlayer}
    \end{minipage}
\end{figure}

Development of language in children takes place in stages and often involves a combination of gestures, vocalizations and use of adequate vocabulary \cite{shilpashri2020pragmatic}. The acquisition of vocabulary facilitates children’s communication in social contexts (known as social communication) and has been shown to be related to academic achievement \cite{ramsook2020you}. Children often are able to comprehend the words spoken to them (receptive vocabulary) before they are able to use them (expressive vocabulary) \cite{mcdaniel2018predicting}. In addition, successful social communication requires the ability to link the conversation partner’s perspective (Theory of Mind) with the context and the words spoken \cite{chin2000teaching}. Thus, understanding and participating in a social conversation requires one to perceive and comprehend the spoken sounds while considering the conversation partner’s perspective (Theory of Mind) and responding using appropriate words.

There is rich literature reporting difficulties in language development of children with Autism \cite{bal2020predictors,saul2020expressive}. Previous research indicates that children with Autism often face difficulties in vocabulary acquisition that, in turn, can have an impact on their social communication \cite{hart2023trajectories}. For teaching vocabulary to children, picture books are often used \cite{ hashemifardnia2018effect} and have been found to be effective in facilitating vocabulary learning in typically developing children \cite{allen2015ipads}. 

In comparison to use of traditional picture books, teaching vocabulary using digital tools might have several advantages for children with Autism. For instance, digital tools can offer a more structured and predictable environment, thereby allowing repetition \cite{guneysu2024enhancing}. In addition, it is easier to provide effective individualized feedback \cite{guneysu2024enhancing}. Various researchers have explored the use of technology-assisted vocabulary learning for children with Autism, e.g., \cite{khowaja2019serious,ganz2015impact,chebli2017generalization}. However, none of these past studies have incorporated turn-taking to facilitate learning and use of vocabulary similar to that used in real-world social contexts.

To address this gap, we propose the design of a cooperative (two-player) vocabulary learning game called CoVoL. By combining turn-taking with interactive elements of a digital picture book, children can engage in game-based vocabulary learning useful for real-world social communication scenarios. We believe such an approach will not only facilitate learning and retention of vocabulary, but also improve Theory of Mind in children with Autism by enhancing their understanding of their conversational partner’s perspective. This might help children with Autism develop social communication skills, improve their academic performance, and boost their overall confidence and self-esteem.

\section{Related Work}
In the following, we discuss related work on vocabulary development in children with Autism, as well as work on technological interventions to address existing issues in this group.

\subsection{Vocabulary Development in Children with Autism}
Up to 50\% of children with Autism demonstrate delay in language development~\cite{smith2007predictors} that can be attributed to difficulties in various sub-components of language \cite{shilpashri2020pragmatic}. These difficulties can range from complete absence of verbal ability to difficulties in the use and comprehension of vocabulary \cite{saldana2023atypical}. Most vocabulary assessments focus on assessing both expressive and receptive vocabulary \cite{hart2023trajectories}, both of which children with Autism might experience difficulties with \cite{hart2023trajectories}. Some studies report atypical expressive-receptive vocabulary development in children with Autism with expressive vocabulary being larger than receptive vocabulary. 
Hart et al.~\cite{hart2023trajectories} find that children with Autism perform poorer than their typically developing counterparts on various aspects of vocabulary starting from 24 months of age. Further, it has been shown that children with Autism are more likely to make semantic errors in addition to being slower in responding to vocabulary tasks \cite{saldana2023atypical}. 
However, despite difficulties in vocabulary acquisition, there seem to be individual trajectories in vocabulary learning for children with Autism \cite{hart2023trajectories}. This necessitates the use of individualized vocabulary learning paradigms, as some children may need more intensive vocabulary based programs than others \cite{hart2023trajectories}.
Such individualized help might be supported by technology.

\subsection{Technology-assisted Vocabulary Learning for Children with Autism}
Given the benefits associated with technology-based interventions for children with Autism \cite{grynszpan2014innovative} such as increased attention span and motivation, feasibility of providing automated practice and feedback, possibility of integrating text, sound and images in parallel \cite{urrea2024use}, many researchers in the past have explored the use of technology for vocabulary learning. Specifically, it has been shown that often children with Autism learn from visual media, which can be easily integrated with the help of technology \cite{urrea2024use}. Previous work has also explored the use of computer-mediated verbal instructions to teach vocabulary to children with Autism \cite{khowaja2019serious}. Ganz et al. studied the efficacy of a tablet computer-based augmentative and alternative communication system with voice output on the receptive-language identification of target words \cite{ganz2015impact}, showing slight improvements. 
Chebli et al. used an application to teach one-word receptive identification to children with Autism \cite{chebli2017generalization}. In their application, an automated voice was used to name the concept. This was followed by the child choosing the image related to the named concept. Novack et al. used a mobile application that included modified Discrete-Trial Training procedures and other behavioral principals of Applied Behavior Analysis (ABA) to teach receptive vocabulary to children with Autism \cite{novack2019evaluation}. 

However, to the best of our knowledge, no study incorporates peer-mediated learning through turn-taking, which can be effective in enhancing social communication in children with Autism \cite{zhang2022effectiveness}.
Additionally, peer-mediated learning might help children in understanding their conversation partner's perspective, i.e., Theory of Mind \cite{chin2000teaching}. 

\section{Method}
To implement and evaluate CoVoL, we follow the Design Science Research (DSR) method \cite{brocke20}.
In DSR, an artifact is designed and evaluated in an iterative problem-solving process -- CoVoL in our case.
Each iteration ends with an evaluation step, which leads to novel insights for the next iteration.
At the time of submission, we completed the second iteration of the DSR cycle, resulting in a functional prototype of CoVoL for a single player.
In the next step, we are extending this to the envisioned two-player game.

\subsection{Iteration 1: Designing Activity Flows}
In the first iteration, we focused on gaining initial understanding of important design principles prior to implementing a CoVoL prototype.
We interviewed three researchers focusing on neurodevelopmental disorders and ABA Verbal Behavior (VB) Therapy based in Poland using a semi-structured interview guide.
After providing our initial ideas regarding the features of CoVoL, the experts provided recommendations on activity flows and scenarios suitable for vocabulary learning for children with Autism.

\subsection{Iteration 2: Proof of Concept}
The second iteration then shifted focus on developing a proof of concept for a single player mode and further investigating core features, so that shifting to the intended two-player mode would be a simple additional implementation step.
In parallel, we developed mechanisms to record technical aspects such as word recognition latency and word error rate.
To evaluate the second iteration, we tested two English Vosk models with 118 audio samples the first author pre-recorded himself.
The 128MB English lgraph model showed an acceptable performance (at over 85\% accuracy), but had a delay of up to 5 seconds on an 2013 MacBook Pro.
The 40MB small English model exhibited only around 50\% accuracy for the same audio samples.
This showed that the larger model is necessary for our use case, and a hosted version might be necessary for slower computers.
%
%
Additionally, we conducted further interviews with two of the experts from the first iteration and, additionally, with five psychologists or speech-language therapists with clinical experience supporting language development in children with Autism.

\subsection{Evaluation Results: Expert Feedback}
Based on the input from the expert interviews in Iteration 1 and 2, we elicited various important requirements for CoVoL.

In general, they all stressed the importance of having a free tool for vocabulary learning, ideally with \textbf{multi-language support}.
In particular, they reported that existing apps are often too expensive for parents to use them at home, and they usually only support English.
They also expressed willingness to try out CoVoL in their settings.

The experts supported the implementation of \textbf{tacting}, i.e., to tact or label objects~\cite{skinner1957verbal}, as it is the easiest intervention, but crucial in everyday life.
They also provided insights into a typical flow, presenting a pictogram every 3 seconds in the beginning, up to 55 pictograms per minute.

In terms of prompts, one expert highlighted that the \textbf{same pictogram might be used for different prompts}, not only teaching to label an object, but also attributes such as shape, color or category. This allows children to understand different dimensions.

Rewards were highlighted by several experts.
First, they stressed that motivation in children varies substantially, and requires customization options, e.g., to \textbf{replace default reward screens}.
The timing and frequency of rewards was also stressed in this context. 
Children get used to overly frequent or fast rewards in electronic applications, thus making it face-to-face human interaction harder.
As a consequence, if implemented wrongly, electronic vocabulary learning tools can actually hurt children using them, as they cannot be used together with standard and evidence-based interventions.
This highlights the importance of implementing \textbf{``human like'' and adjustable response times and configuring the frequency of rewards}.

The traditional ABA VB tacting protocol assumes one therapist that prompts a child to label an object., without considering cooperative learning.
While the interviewed experts all come from an ABA VB background and were excited about a free game that implements the traditional protocol, they also stressed that the cooperative game idea could be promising.
In particular, one expert mentioned that, currently, success in therapy sessions does not always generalize to improved communication outside the therapy.
\textbf{Cooperation could motivate children to learn social communication} and, thus, allow for a more generalizable learning experience.
They highlighted that the two-player game mode models learning in real life, which often takes place through imitation of others' behaviors.

\section{CoVoL -- A Cooperative Vocabulary Learning Game}
\label{sec:covol}
In the following, we outline the features and technical aspects of CoVoL.

\begin{figure}
    \centering
    \includegraphics[width=0.2\textwidth]{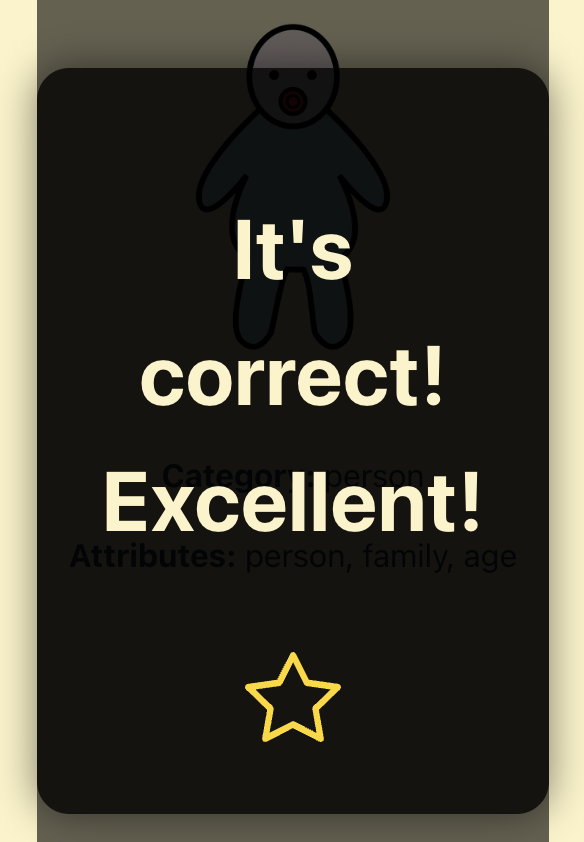}
    \caption{Reward screen for a correctly-labeled object.}
  \label{fig:rewardScreen}
\end{figure}

The underlying purpose of CoVoL is to teach children tacting.
Improving tacting skills in children with Autism is important, as it can lead to less repetitive and stereotypical language \cite{karmali2005reducing}.
Following ABA recommendations, in single player mode, we show players a pictogram of an object and prompt them to label this object verbally (see Figure~\ref{fig:promptExample}).
If the player labels the object correctly, a reward screen is shown (see Figure~\ref{fig:rewardScreen}).
Afterwards, the process is repeated with the next object.
This process continues for a defined amount of objects.

In two-player mode, two players take turns labeling objects in the same way as for single player mode.
Visually, we display two player avatars on the left-hand side and right-hand side of the screen (see Figure~\ref{fig:promptExampleTwoPlayer}).
We highlight the avatar who's turn it is, while the other avatar is grayed out.
As in single player mode, a reward screen is shown if the object is labeled correctly.
Afterwards, it is the other player's turn.
In the two player mode, a child learning to speak could be paired with a therapist, a parent/guardian, or another child, potentially depending on their speaking level.
In the current prototype of CoVoL, prompts are only displayed textually.
Similarly, the players are displayed as ``Player 1'' and ``Player 2'' only.
In terms of tact recognition, we recognize several synonyms for each object.
For instance, the apple prompt in Figure~\ref{fig:promptExampleTwoPlayer} would recognize if the user says ``apple'', ``apples'' or ``fruit''.

CoVoL uses a simple three-tier architecture.
The frontend is implemented in TypeScript using the Vite framework\footnote{\url{https://vite.dev/}}.
The backend is implemented in Go using the Fiber framework\footnote{\url{https://gofiber.io/}} and Vosk\footnote{\url{https://alphacephei.com/vosk/}} as a voice recognition library.
Communication between the frontend and backend takes place through WebSockets.
Prompts are stored in JSON files so that non-technical users could easily extend or modify the objects used for tacting.
Similarly, object images are stored as regular image files.
We use the Augmentative and alternative communication (AAC) image set provided\footnote{The pictographic symbols used are the property of the Government of Aragón and have been created by Sergio Palao for ARASAAC (\url{http://www.arasaac.org}), that distributes them under Creative Commons License BY-NC-SA.} by the Aragonese Center of Augmentative and Alternative Communication (ARASAAC).
%


\subsection{Planned Features}
Based on the obtained expert feedback, we envision various features for CoVoL in addition to the currently implemented aspects.
First, we will enable the use of audio prompts and feedback instead (or in addition to) the textual prompts.
Additionally, we will allow players to define their own names instead of ``Player 1'' and ``Player 2'', as well as choose an avatar among several options instead of the icons currently displayed (see Figure~\ref{fig:promptExampleTwoPlayer}.

Supporting therapists and parents, we will expose various configuration options to the end users.
This includes setting a \textbf{minimum delay} for the reward, so that children get used to waiting a short while until they receive feedback.
Furthermore, we will allow \textbf{changing the reward screen}, to tailor it to children's preferences.
For instance, the star could be replaced with another icon.
We will implement \textbf{multi-language support} in CoVoL.
Since Vosk already supports a multitude of languages, this change requires only added support to translate textual prompts and a configuration option to change the language.
However, we are also designing the underlying architecture so that Vosk could be exchanged for other language models.
Specifically, we are investigating to use an Icelandic language model, tailored to the local context.

Going beyond the basic tacting intervention, we also plan to support different labeling flows.
That is, prompts could be changed to ask for attributes of the displayed objects.
For example, a player could identify the apple in Figure~\ref{fig:promptExampleTwoPlayer} as ``red'' or ``round''.
The data structure used to define the various objects already supports adding attributes, as well as alternative object descriptions (such as ``fruit'' for the apple).
We also envision to support flows that cannot be emulated by a human therapist, such as playing sounds and prompting children to label these.

\subsection{Planned Evaluation}
In this prototype stage, we only evaluated CoVoL through ten expert interviews.
While the author team has experience conducting research in neurodiversity, including children with Autism, and have family members with Autism, any form of evaluation or testing with the intended target group is currently missing.
To address this, we are planning to evaluate CoVoL in three scenarios, subject to ethics approval which we are preparing at the moment.
First, we will pair one child each with a therapist and evaluate the two-player mode.
That is, the therapist and child will take turns labeling objects.
Second, we will perform the same kind of evaluation, but pairing a child with one of their parents/guardians instead.
In particular, we are interested in how CoVoL can help parents in supporting their children, thus empowering them and reducing the need for therapy sessions.
Third, we aim to conduct an exploratory evaluation in which we will pair two children.
We will explore both pairs where both children are diagnosed with ASD and pairs where only one child is diagnosed with ASD.
Therapists or parents/guardians will be included as observers in this scenario.
We will observe and qualitatively analyze all interactions, as well as following up with interviews with therapists/guardians.
These three evaluation scenarios also mirror the intended usage of CoVoL, depending on the child's speaking aptitude.


\subsection{Limitations}
We are currently aware of three limitations in CoVoL.
First, stuttering can lead to difficulties in detecting words.
As stuttering is more common in people with Autism than without \cite{pirinen23}, this could pose an important barrier to using CoVoL in the intended target group.
We will evaluate the extent of this limitation in future work.
Second, we currently detect labels on a word level, i.e., whether the correct word has been uttered by the player.
This means that we cannot detect negations, as a sentence like ``This is not an apple.'' would lead Vosk to detect that ``apple'' has been uttered, thus determining that the object has been correctly labeled.
We believe this is not a strong limitation, as we focus on fundamental vocabulary learning only.
Finally, we evaluated the quality of Vosk's word detection with a single person.
Further evaluations are necessary to better understand how various dialects or other differences in speaking affect detection quality.


\section{Conclusion}
We present ongoing work on CoVoL, a two-player game for collaborative vocabulary learning in children with Autism.
We designed a prototype in close interaction with researchers in neurodevelopmental disorders and therapists with clinical experience working with children with Autism.
The current prototype supports single player mode mirroring the traditional interaction with a therapist.
In the next step, we will extend this to a two-player setup with turn taking and evaluate this setup with children.
The expert interviews provide rich insights into critical requirements for the design of CoVoL, such as configurable delays to mimic real-life delays in social interaction.


\bibliographystyle{plain}


\end{document}